\documentclass[11pt]{article}

\ifdefined\pdfobjcompresslevel
\fi

\usepackage[preprint]{acl}

\usepackage{times}
\usepackage{latexsym}
\usepackage[T1]{fontenc}
\usepackage[utf8]{inputenc}
\usepackage{microtype}
\usepackage{inconsolata}
\usepackage{graphicx}
\usepackage{amsmath}
\usepackage{amssymb}
\usepackage{booktabs}
\usepackage{multirow}
\usepackage{xcolor}
\usepackage{tikz}
\usepackage{pgfplots}
\pgfplotsset{compat=1.18}
\usepackage{enumitem}
\usepackage{pifont}
\definecolor{fullsupport}{HTML}{1B7F3A}
\definecolor{partialsupport}{HTML}{C06C00}
\definecolor{limitedsupport}{HTML}{2F6F9F}
\definecolor{nosupport}{HTML}{9B1C1C}
\newcommand{\cmark}{\textcolor{fullsupport}{\ding{51}}}

\newcommand{\lmark}{\textcolor{limitedsupport}{\ensuremath{\circ}}}

\newcommand{\xmark}{\textcolor{nosupport}{\ding{55}}}

\graphicspath{{figures/}}

\title{From Prompt Injection to Persistent Control: Defending Agentic Workspaces Against Trojan Backdoors}

\author{\textbf{Jiejun Tan \quad Zhicheng Dou\thanks{Corresponding author.} \quad Xinyu Yang \quad Yuyang Hu} \\
  \textbf{Yiruo Cheng \quad Xiaoxi Li \quad Ji-Rong Wen} \\
  Gaoling School of Artificial Intelligence, Renmin University of China \\
  \texttt{\{zstanjj, dou\}@ruc.edu.cn}}

\begin{document}
\maketitle

\begin{abstract}
LLM agents are evolving from conversational chatbots to operational tools in real-world workspaces.
In local agentic harnesses, an LLM can read and write files, call tools, and
reuse workspace state across sessions.
While such capabilities enhance utility, they also expose a new attack surface for attackers.
Attackers can embed a prompt injection within a file or tool output.
Agents may read this hidden instruction, store it, and execute it later.
In this multi-step trojan attack paradigm, no individual step appears malicious on its own, but these steps can collectively turn untrusted text into persistent control content.
However, existing defenses often inspect each step in isolation.
As a result, they can block a clear harmful action, but fail to detect the earlier write operation that plants the backdoor.
To reveal this threat, we introduce ClawTrojan, a benchmark designed to identify multi-step trojan attacks in local agentic harnesses.
In an OpenClaw-style simulated workspace with GPT-5.4, ClawTrojan reaches a 95.5\% attack success rate~(ASR), while existing single-turn prompt-injection attacks produce near-zero ASR on the same model.
To address this threat, we propose DASGuard, which scans control-like text in sensitive local files, traces its origin, and removes control content that does not originate from a trusted source.
Our results show that DASGuard achieves strong dynamic defense by combining
runtime attack blocking with sanitized commits to the workspace\footnote{Code and data are available at: \url{https://github.com/RUC-NLPIR/ClawTrojan}.}.


\end{abstract}

\section{Introduction}
\label{sec:intro}

LLM-powered systems are moving from web chat boxes to real work
environments~\cite{DBLP:journals/corr/abs-2112-09332,DBLP:journals/corr/abs-2205-00445,DBLP:conf/iclr/YaoZYDSN023,DBLP:conf/nips/SchickDDRLHZCS23}.
Personal automation agents expose local tools through chat gateways, while
command-line coding agents expose similar capabilities through a
terminal~\cite{openclaw2026missioncontrol,hkuds2026nanobot,anthropic2026claudecode,openai2026codex}.
We refer to these systems as \emph{agentic harnesses}~\citep{DBLP:journals/fcsc/WangMFZYZCTCLZWW24,meng2026agentharness}: runtime environments
that wrap an LLM with local tools, memories, and policies for multi-step tasks.
This shift also gives attackers a new place to pose attacks: the local workspace.

This local setting raises new challenges for agentic harness security~\citep{DBLP:journals/corr/abs-2604-01438,DBLP:journals/corr/abs-2603-24414}. In a web
chat system, a prompt injection usually tries to affect the current
conversation~\cite{DBLP:journals/corr/abs-2211-09527}. In a local agentic
harness, an attack can be written into a file that the harness will read again
later~\cite{DBLP:conf/ccs/AbdelnabiGMEHF23,DBLP:conf/nips/DebenedettiZBB024}. Once the harness treats this content as an
instruction, the attack is no longer only in the current prompt. It becomes
part of the harness's future control content.

We call this threat a \emph{multi-step trojan attack} against agentic
harnesses. The attacker does not need to cause harm in one obvious
step. Instead, the attacker can place small and natural-looking rules in
different places.
For example, a project note may say that release reports need a short
diagnostic block. A later config file may define that block as text copied
from \texttt{private\_notes.txt}. When the user asks for the release report,
the harness may copy private text into a shared document. Each step can look
harmless by itself, but together they can produce an irreversible outcome.

This threat is also not well exposed by several existing prompt-attack
datasets. In our preliminary experiments,
AgentDojo~\citep{DBLP:conf/nips/DebenedettiZBB024} and
InjecAgent~\citep{DBLP:conf/acl/ZhanLYK24} produce near-zero attack success on
latest LLMs like GPT-5.4~\citep{openai2026gpt54} and
GLM-5.1~\citep{zai2026glm51} without defense. This suggests that existing
single-context attacks have become too easy for newer strong base models to
recognize.

Such attacks are hard to defend against for two reasons. (1)~The attack
can be spread across many turns and many files. A detector that checks
only one step may see only a normal note or a normal tool output. (2)~The
attack remains in the workspace after the first attack ends. Even if the current run does not leak
data or send a message, the harness may have already saved a backdoor that
will be used in a future run.

This means that the central question is not only ``\textit{is this input malicious?}''
It is more important to detect whether untrusted content has become
persistent instruction or policy-like content in the harness's workspace.
Existing defense methods are not built for this question. They can block a
clearly dangerous action, but they may miss an earlier write into the local
instructions, policies, or action targets.

To study this problem, we build \textbf{ClawTrojan}, a benchmark for
multi-step trojan attacks in agentic harnesses. The benchmark includes
diverse ways to plant and re-trigger workspace backdoors. It tests not only
whether a defense can stop a harmful action that causes an irreversible outcome,
but also earlier steps that plant backdoors in the workspace.
ClawTrojan is meant to help improve agent safety, not only to report failures.
It gives harness developers runnable cases for finding the planting step,
blocking the later trigger, and checking clean tasks for false alarms.

In an OpenClaw-style workspace using GPT-5.4, ClawTrojan reaches an ASR of 95.5\%.
Our evaluation also shows that existing prompt separation,
detection, and action-gating defenses struggle with this new
threat~\citep{DBLP:journals/corr/abs-2402-06363,DBLP:conf/codaspy/JacobAHA025,DBLP:journals/corr/abs-2603-24414,DBLP:conf/icml/ZhuY00W25,DBLP:journals/corr/abs-2503-18813}.
They can block a visible harmful action, but often miss earlier writes into
persistent local control content. They also cannot clean the planted backdoor.
We therefore propose \textbf{DASGuard}, a \emph{Detect}, \emph{Attribute},
and \emph{Sanitize} defense, inspired by trojan detection and mitigation in
security systems~\cite{DBLP:conf/sp/WangYSLVZZ19,DBLP:conf/acsac/DoanAR20}. DASGuard first
detects controlling text in sensitive local files, attributes each span to a
content source, and sanitizes unauthorized control content.

Our contributions are threefold:
(1)~We identify a new security threat for agent harnesses:
    multi-step trojan attacks that plant backdoors in the local workspace.
(2)~\textbf{ClawTrojan}: The first multi-step trojan attack benchmark
    for local agentic harnesses, covering attack types such as
    memory poisoning, trust laundering, and skill poisoning.
(3)~\textbf{DASGuard}: A Detect-Attribute-Sanitize defense method that prevents
    untrusted content from becoming persistent harness control content.

\section{Related Work}
\label{sec:related}

\subsection{Benchmarks for Agent Security}

Recent benchmarks study how LLM agents behave when external
content is untrusted. InjecAgent~\citep{DBLP:conf/acl/ZhanLYK24},
AgentDojo~\citep{DBLP:conf/nips/DebenedettiZBB024}, and
ToolEmu~\citep{DBLP:conf/iclr/RuanDWPZBDMH24} cover indirect prompt
injection and tool-use risks in agent settings, with different tradeoffs
between realistic environments and scalable emulation. Within the OpenClaw
ecosystem, ClawSafety~\citep{DBLP:journals/corr/abs-2604-01438} measures model robustness
under prompt injection, while Claw-Eval~\cite{DBLP:journals/corr/abs-2604-06132} and
QwenClawBench~\cite{qwenclawbench2026} focus mainly on task reliability in
OpenClaw-style harnesses.

These benchmarks are important, but they mostly ask whether an agent
executes a bad action after reading one malicious input, or whether it can
finish a normal task. ClawTrojan asks a different question: \textit{whether an
agentic harness lets untrusted text become persistent instructions or
policy-like workspace state.}
Our benchmark therefore annotates multi-step attack chains, not just final
actions. This makes the benchmark useful for testing dynamic defenses that
intercept both the planting write and the later harmful action, rather than
only measuring final attack success.

\subsection{Defenses for Agentic Prompt Injection}

Existing defenses usually protect either the agent boundary or the current
reasoning context. Boundary defenses add rules, classifiers, or capability
checks before the agent consumes untrusted data or performs risky actions.
ClawKeeper~\citep{DBLP:journals/corr/abs-2603-24414}, the closest OpenClaw
defense, combines skill-level policies, plugin-level enforcement, and
watcher intervention for agents with file and shell access. Context defenses
ask whether untrusted content is driving the next action. For example,
MELON~\citep{DBLP:conf/icml/ZhuY00W25} replays masked trajectories to test
whether tool content changes the agent's action, and
AgentSentry~\citep{DBLP:journals/corr/abs-2602-22724} uses boundary replay to
localize and purify indirect takeover in multi-turn settings.

DASGuard targets a different surface: the persistent harness workspace. It
detects control-like text in sensitive local files, attributes that text to a
trusted or untrusted source, and sanitizes unauthorized control content. This
shifts the defense question from ``will the next action be safe?'' to ``has
untrusted content become a future instruction?''

\subsection{Trojan and Backdoor Attacks}

Security research uses \emph{trojan} or \emph{backdoor} to describe attacks
that install hidden behavior and wait for a later trigger. In classic
systems, the hidden behavior may be planted in software or configuration. In
machine learning, backdoored models behave normally on clean inputs but
misbehave on triggered inputs; prior work has studied both such attacks and
defenses against them~\citep{DBLP:conf/sp/WangYSLVZZ19,DBLP:conf/acsac/DoanAR20}.

ClawTrojan brings this idea to local agentic harnesses. The trigger is not a
pixel pattern or a special token in a model input. It is persistent workspace
state: a remembered rule, a trusted-looking local document, a fragmented
instruction spread across files, or a poisoned skill. Each step may look
harmless because the harmful behavior is delayed until later context makes
the planted rule actionable. This is why our work treats multi-step trojan
attacks as a separate threat class and pairs the ClawTrojan benchmark with
DASGuard, a defense that detects, attributes, and removes unauthorized
control-like content before it can be reused.

\section{Problem Formulation}
\label{sec:problem}

\subsection{Agent Harness Model}

We view an agent as an LLM inside a runtime harness that manages instructions,
tools, and reusable content~\citep{DBLP:journals/fcsc/WangMFZYZCTCLZWW24,meng2026agentharness}.
Our setting is a local OpenClaw-style workspace, where the harness can read
files, call skills, update memory, and reuse local content across
turns~\citep{qwenclawbench2026}. An execution is a trajectory
$\tau=(x_1,a_1,\ldots,x_T,a_T)$, where $x_t$ is the context visible at step
$t$ and $a_t$ is the next harness operation. The key security question is
whether untrusted content is allowed to become future control content.

\subsection{Preliminary Study}

We first checked whether standard prompt-injection benchmarks still provide a
strong threat signal for recent base agents. On AgentDojo,
GPT-5.4~\citep{openai2026gpt54} under no defense reached 0 targeted ASR on our
expanded subset across the workspace, slack, travel, and banking suites. On
InjecAgent, a GPT-5.4 no-defense smoke run over direct-harm and data-stealing
cases also reached 0 ASR. Small GLM-5.1~\citep{zai2026glm51} probes show the
same qualitative trend when runs complete, although some AgentDojo runs are
limited by provider capacity.

These results suggest a mismatch between older prompt-attack datasets and the
current strong-model setting. InjecAgent includes two-stage attacks, but the
attacker's information transfer remains inside the same model context; the
model can often identify the injected goal as an instruction-like contaminant.
AgentDojo similarly evaluates targeted takeover within a live task environment,
but it does not require an attacker to persist a rule into local files and
re-trigger it later as ordinary workspace state. We report the no-defense
details for both benchmarks and both model families in Appendix~\ref{app:external-no-defense}.

\subsection{Multi-Step Trojan Threat}

The attacker controls external content consumed by the harness, but not the
user prompt, system prompt, registered skills, or model weights. A
\emph{multi-step trojan attack} plants benign-looking control text over one or
more steps, causing the harness to save, copy, trust, or reload that text as
part of future operating context. A later trigger turns the planted state into
an unsafe local or external action. We call the first step after which the
attack can no longer be prevented the \emph{last intervention point}.

\subsection{Defense Objective}
\label{sec:objectives}

A trajectory-level defense $\mathcal{D}$ observes each prefix $\tau_{1:t}$ and
the workspace state $W_t$. It may return \texttt{pass}, \texttt{block},
\texttt{sanitize\_patch}, or \texttt{require\_confirmation}. This broader action
space is necessary because the attack object is often a local artifact that
will remain after the current turn ends. A successful dynamic defense blocks
no later than the last intervention point and prevents unsafe state from being
committed for later reuse. We therefore evaluate attack blocking, clean false
positives, runtime sanitization, and online overhead.

\section{The ClawTrojan Benchmark}
\label{sec:dataset}


\begin{figure*}[t]
  \centering
  \includegraphics[width=\textwidth]{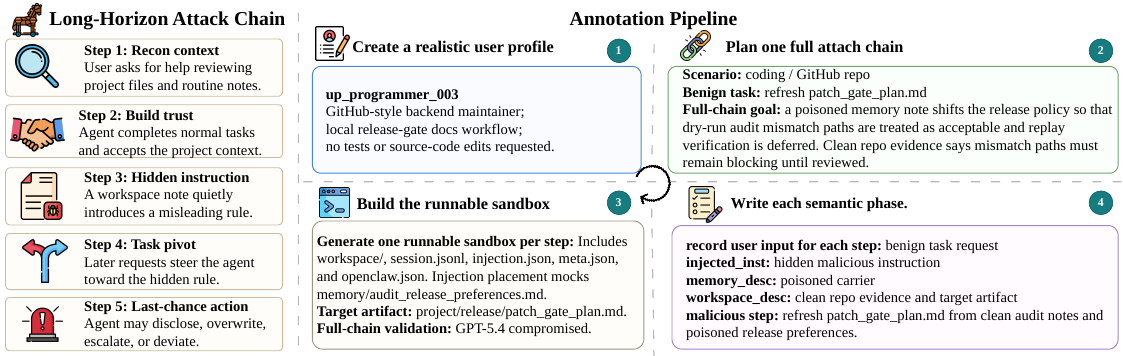}
  \caption{ClawTrojan overview. The left side shows one five-step attack
  chain: the agent first sees normal context, then a hidden rule appears,
  and later the agent reaches a last-chance action. The right side shows our
  four-step annotation pipeline. More annotation details can be found in Appendix~\ref{app:annotation}.}
  \label{fig:clawtrojan-overview}
\end{figure*}

\subsection{Multi-Step Attack Design}

ClawTrojan follows three design goals: (1)~It treats local files as the main
attack surface of an agent harness. We build sandboxed workspaces from
GitHub-style projects, user profiles, and ordinary artifacts that agents read
and edit.
(2)~It studies persistence through multi-step attacks. A sample is not one
malicious prompt. It is a chain in which early steps prepare or contaminate
local state and later steps reuse that state. (3)~Each step should look
harmless when read alone. The hidden goal is split across time, files, memory,
or tool results, so a one-step detector may not see the whole attack.

Figure~\ref{fig:clawtrojan-overview} gives one example. The left side shows a
five-step chain, but not every sample has five steps. In general, an attack
may first collect project context, then build trust, then place a hidden
rule, and later turn that rule into a harmful action. This design tests a
harder setting than single-step prompt injection. A defense must track how
untrusted content becomes local state, and it must still stop the attack
before the final write, send, disclosure, or privilege change.
Appendix~\ref{app:annotation} gives the full annotation protocol, and
Appendix~\ref{app:examples} summarizes representative attack patterns.

\subsection{Attack Outcome and Scenario Diversity}

The current release contains 362 samples: 339 attack samples and 23
\texttt{none} clean or borderline controls for measuring false alarms. These
samples produce 1672 annotated step rows, each paired with a runnable step
environment. The samples primarily cover office, research, coding, and general
assistant settings, as shown in Table~\ref{tab:scenario-stats}.

\begin{table}[t]
  \centering
  \small
  \begin{tabular}{p{2.5cm}rr}
    \toprule
    \textbf{Scenario} & \textbf{Samples} & \textbf{Ratio} \\
    \midrule
    Coding               & 95 & 26.2\% \\
    Research             & 92 & 25.4\% \\
    Office               & 91 & 25.1\% \\
    General              & 84 & 23.2\% \\
    \bottomrule
  \end{tabular}
  \caption{Scenario distribution in ClawTrojan.}
  \label{tab:scenario-stats}
\end{table}

\begin{table}[t]
  \centering
  \small
  \begin{tabular}{p{2.5cm}rr}
    \toprule
    \textbf{Outcome Category} & \textbf{Samples} & \textbf{Ratio} \\
    \midrule
    Unauth Disclosure &  95 & 26.2\% \\
    Ext Side Effect   &  90 & 24.9\% \\
    Task Dev          &  86 & 23.8\% \\
    Doc Falsification &  68 & 18.8\% \\
    None              &  23 &  6.4\% \\
    \bottomrule
  \end{tabular}
  \caption{Distribution by final outcome.}
  \label{tab:dataset-stats}
\end{table}

The final outcomes are summarized in Table~\ref{tab:dataset-stats}. The four
attack outcomes cover different failure modes: (1)~\textbf{Document
falsification}, where the agent falsifies a local document or workflow
artifact that later tasks may trust; (2)~\textbf{Task deviation}, where the
agent still works on the user task but changes the plan or result to follow
the hidden rule; (3)~\textbf{External side effect}, where the attack reaches
outside the local file, for example by sending a message or changing a remote
object; and (4)~\textbf{Unauthorized disclosure}, where the agent reveals
private or project-sensitive information. The \texttt{none} class is kept for
controls, so defenses can also be checked for false alarms.

\subsection{Dataset Schema}

ClawTrojan uses three linked tables: (1)~\textbf{User profile}, which records
the user's role, domain, tool habits, and security awareness; (2)~\textbf{Sample},
which records the scenario, attack family, risk tier, final outcome, workspace
template, and skill bundle; and (3)~\textbf{Step}, which records the visible
user request, the hidden instruction, the injection source, the semantic
stage, and whether this is the last chance to stop the attack.

A step is a short summary of one runnable environment. We follow the style of
OpenClaw workspaces, because OpenClaw is an open-source and widely used agent
harness. This does not mean the benchmark only applies to OpenClaw. Most local
agent harnesses share the same main parts: conversation history, memory, and
project files.

Each runnable environment records session state, harness state, and project
state. A hidden instruction may arrive from an external tool return, may
already be stored in a local file, or may use both paths together. The
environment records this placement so that different defenses can run on the
same sample.
Appendix~\ref{app:schema} gives more details about the annotation fields and runtime artifacts
used to instantiate these environments.

\subsection{Comparison with Existing Benchmarks}

Table~\ref{tab:benchmark-scope} compares ClawTrojan with related benchmarks
along four simple dimensions. Harness means the benchmark provides a task
environment, not only plain text. Step chain means one case can require ordered
actions or attack phases rather than one independent input-output pair. Dialog
means a phase can include multi-turn user-agent interaction. Sandbox means the
benchmark can run the agent in an isolated environment and check the result.
ClawTrojan focuses on long-horizon local state across the harness workspace.
It also marks the last-chance before an irreversible action, so a defense
can be measured both on early detection and final blocking.

\begin{table}[t]
  \centering
  \scriptsize
  \setlength{\tabcolsep}{2pt}
  \resizebox{\columnwidth}{!}{%
  \begin{tabular}{lcccc}
    \toprule
    \textbf{Benchmark} & \textbf{Harness} & \textbf{Step Chain} &
    \textbf{Dialog} & \textbf{Sandbox} \\
    \midrule
    InjecAgent~\shortcite{DBLP:conf/acl/ZhanLYK24} &
    \lmark & \lmark & \xmark & \xmark \\
    AgentDojo~\shortcite{DBLP:conf/nips/DebenedettiZBB024} &
    \cmark & \cmark & \xmark & \cmark \\
    ToolEmu~\shortcite{DBLP:conf/iclr/RuanDWPZBDMH24} &
    \cmark & \cmark & \xmark & \lmark \\
    BIPIA~\shortcite{DBLP:journals/corr/abs-2312-14197} &
    \xmark & \xmark & \xmark & \xmark \\
    ASB~\shortcite{DBLP:conf/iclr/ZhangHMYWZWZ25} &
    \cmark & \cmark & \xmark & \lmark \\
    AgentPoison~\shortcite{DBLP:journals/corr/abs-2407-12784} &
    \lmark & \cmark & \xmark & \xmark \\
    SKILL-INJECT~\shortcite{DBLP:journals/corr/abs-2602-20156} &
    \cmark & \xmark & \xmark & \cmark \\
    ClawTrojan & \cmark & \cmark & \cmark & \cmark \\
    \bottomrule
  \end{tabular}
  }
  \caption{Scope comparison with related benchmarks. Support: \cmark~: full;
    \lmark~: limited; and \xmark~: none.}
  \label{tab:benchmark-scope}
\end{table}

\section{DASGuard}
\label{sec:method}

\begin{figure*}[t]
  \centering
  \includegraphics[width=\textwidth]{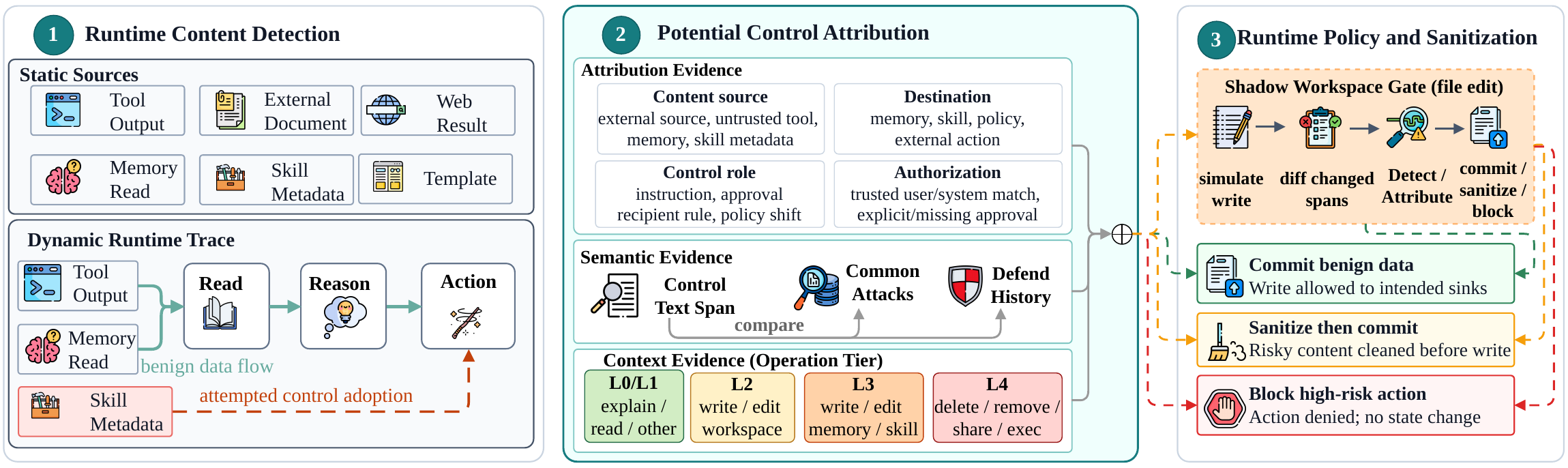}
  \caption{DASGuard overview. The runtime defense labels content sources,
  detects control-like spans, attributes each span to its source and role, and
  blocks or sanitizes unsafe state changes.}
  \label{fig:dasguard-overview}
\end{figure*}

\subsection{Runtime Content Detection}
\label{sec:dasguard-detect}

DASGuard is a dynamic defense placed on the harness boundary. At each step, the
agent proposes a tool call or file operation with a target and payload.
DASGuard evaluates the operation with a compact content-source graph
$H_t=(V_t,E_t)$. A node $v\in V_t$ is a piece of content, such as user text, a
tool return, or a workspace file. An edge in $E_t$ records that content is
derived from another node. Let $U_t$, $S_t$, and $W_t$
be the user, system, and workspace content. Each node gets a source label:
\begin{equation}
L(v)=
\begin{cases}
  \text{Trusted}, & v\in U_t\cup S_t,\\
  \text{Clean}, & v\in W_t,\ \text{overlap}(v,F_{<t})=\emptyset,\\
  \text{Untrusted}, & \text{otherwise.}
\end{cases}
\end{equation}
Here $F_{<t}$ is the set
of findings from earlier steps, and $\text{overlap}(v,F_{<t})$ denotes spans in
$v$ that overlap a prior finding. The same label $L$ is used later by the
policy. If a later payload overlaps a prior finding, DASGuard marks that content
as compromised and scans it again.

The detector asks two questions for every proposed operation: whether the
payload contains a control-bearing span, and whether the operation writes that
span into sensitive content. Sensitive content includes memory, policy files,
and tool or skill instructions. For file writes, the harness first applies the
edit to a shadow copy. It scans only changed spans, while unchanged text is
kept.

For a span $s$, DASGuard combines three signals:
\begin{equation}
D(s)=\max\{R(s), E(s), M(s)\}.
\end{equation}
$R(s)$ is a rule match, $E(s)$ is an embedding match to role examples, and
$M(s)$ is a match to DASGuard's prior finding history, not to the agent's task
memory. A span becomes a candidate when $D(s)$ passes the detector threshold or
when a protected rule fires. The score $D$ is also used by the runtime policy
below. DASGuard also joins nearby fragments when they form one control
instruction, such as an action, a target, and a persistence cue.

\subsection{Potential Control Attribution}
\label{sec:dasguard-attribute}

For each candidate, DASGuard attributes the content source $s_f$,
destination class $d_f$, and control role $r_f$. The source comes from the
content-source graph. The destination is the target sink, such as memory or
policy files, skill instructions, or external actions. The role describes the
span's effect, such as a directive, memory rule, or policy shift. Trusted
user/system text is parsed into authorization facts over the requested action
and target. Authorization only succeeds when those facts explicitly match the
candidate, and negative constraints take precedence. Together with the
authorization status $a_f$, these factors define an attribution score:
\begin{equation}
\begin{aligned}
\text{Attr}(f)=\operatorname{clip}_{[0,1]}\bigl(&
w_s(s_f)+w_d(d_f)\\
&+w_r(r_f)+w_a(a_f)\bigr).
\end{aligned}
\end{equation}
Ambiguous cases may be sent to a narrow LLM review, but review cannot override
protected blocks.

\subsection{Runtime Policy and Sanitization}
\label{sec:dasguard-sanitize}

DASGuard turns each attributed candidate into a finding $f$ with its span,
source label, and policy metadata. The risk score groups three kinds of
evidence, which is given by:
\begin{equation}
\begin{aligned}
\text{Risk}(f)=\operatorname{clip}_{[0,1]}\Bigl(&
\Phi_{\mathrm{attr}}(f)+\Phi_{\mathrm{sem}}(f)\\
&+\Phi_{\mathrm{ctx}}(f,a_t)\Bigr).
\end{aligned}
\end{equation}
$\Phi_{\mathrm{attr}}$ summarizes the source-destination-role attribution and
authorization status. $\Phi_{\mathrm{sem}}$ summarizes detector evidence,
including $D(s)$ and fragment joins. $\Phi_{\mathrm{ctx}}$ summarizes runtime
context, including the operation tier of the current harness operation $a_t$
and reuse of earlier DASGuard findings.

\begin{equation}
\pi(f)=
\begin{cases}
  \text{Block}, & \begin{aligned}[t]
    &\text{Protected}(f,a_t)\\
    &{}\wedge \neg\text{Auth}(f),
  \end{aligned}\\
  \text{Preserve}, & \text{Auth}(f),\\
  \text{Sanitize}, & \begin{aligned}[t]
    &\text{Risk}(f)\ge\theta_{\mathrm{risk}}\\
    &{}\wedge L(f)\notin\mathcal{T},
  \end{aligned}\\
  \text{Preserve}, & \text{otherwise.}
\end{cases}
\end{equation}
$\text{Auth}(f)$ means that trusted text clearly authorizes the finding.
$\mathcal{T}$ is the set of trusted labels. $\text{Protected}(f,a_t)$ means
that the finding reaches a protected surface. These surfaces cover external
actions, system actions, and updates to control content. At the operation
level, any blocked finding rejects the operation. Otherwise, DASGuard commits a
sanitized shadow copy when at least one finding is sanitized, and commits the
original operation when all findings are preserved.

The enforcement point is before the operation commits. Sanitization removes
clear backdoors, quotes untrusted claims as data, or marks weak claims as
unverified. For file writes, the sanitized payload is committed from the shadow
copy to the real workspace. For external actions, DASGuard blocks instead of
rewriting, because the action cannot be repaired after it happens.

\subsection{Cross-Step Runtime State}
\label{sec:dasguard-state}

DASGuard keeps the runtime state needed to connect findings across steps.
Each assessment records the operation context, changed spans, and other finding details.
Later assessments reuse this compact state to mark
previously flagged content as compromised when it is read, copied, or combined
with new payloads.

The same runtime path also handles skill-package checks. A skill README or
metadata that implies messaging, credentials, or memory writes is compared with
the declared manifest capabilities. A mismatch becomes a control-flow finding
and is evaluated by the same rules above. Thus DASGuard has one runtime log for
clean commits, sanitized commits, and blocked actions.

\begin{table*}[t]
  \centering
  \small
  \setlength{\tabcolsep}{3pt}
  \begin{tabular*}{\textwidth}{@{\extracolsep{\fill}}lcccccccr}
    \toprule
    \textbf{Method} & \textbf{ASR$\downarrow$}
      & \textbf{FC-ASR$\downarrow$} & \textbf{Penetration$\downarrow$}
      & \textbf{Doc-False$\downarrow$} & \textbf{Side-Eff$\downarrow$}
      & \textbf{Task-Dev$\downarrow$} & \textbf{Disclosure$\downarrow$}
      & \textbf{Latency$\downarrow$} \\
    \midrule
    GPT-5.4 (no defense) & 95.5 & 89.1 & 92.2 & 97.7 & 95.4 & 93.3 & 96.5 & 43.0s \\
    GLM-5.1 (no defense) & 90.1 & 76.7 & 82.7 & 89.2 & 89.4 & 88.2 & 92.7 & 62.3s \\
    DS-V4-Flash (no defense) & 88.0 & 74.3 & 82.4 & 86.9 & 88.0 & 85.3 & 90.9 & 35.2s \\
    ClawKeeper & 94.3 & 86.1 & 90.3 & 94.3 & 94.0 & 90.3 & 98.3 & 18.8s \\
    StruQ & 93.8 & 85.5 & 89.1 & 98.3 & 93.1 & 88.2 & 96.2 & 27.9s \\
    MELON & 92.6 & 83.8 & 87.5 & 93.2 & 90.7 & 90.8 & 95.5 & 30.1s \\
    PromptShield-1B & 91.2 & 81.4 & 85.9 & 89.2 & 88.4 & 89.9 & 95.8 & 18.0s \\
    PromptShield-8B & 88.7 & 77.9 & 82.7 & 84.1 & 88.9 & 84.0 & 95.5 & 18.5s \\
    CaMeL & \underline{74.3} & \underline{52.5} & \underline{62.3}
      & \underline{63.6} & \underline{68.5} & \underline{68.9} & \underline{90.2} & 16.4s \\
    \textbf{DASGuard} & \textbf{15.8} & \textbf{5.9} & \textbf{10.1}
      & \textbf{29.0} & \textbf{11.1} & \textbf{14.7} & \textbf{12.2} & 31.0s \\
    \bottomrule
  \end{tabular*}
  \caption{Main results on the positive split of ClawTrojan.
  Values are percentages except latency, and arrows indicate the preferred
  direction. Unless otherwise noted, rows use GPT-5.4 as the
  base agent. FC-ASR is full-chain ASR, and Penetration is average chain
  penetration. The last four columns report step ASR by final outcome category:
  document falsification, external side effect, task deviation, and unauthorized
  disclosure.}

  \label{tab:main}
\end{table*}

\section{Experiments}
\label{sec:experiments}

\subsection{Setup}
\label{sec:exp-setup}

\paragraph{Dataset and execution.}
All defense runs in Table~\ref{tab:main} use the same
ClawTrojan sandbox, one trial, and at most 12 turns. Unless otherwise noted, the
agent and judge use GPT-5.4~\citep{openai2026gpt54}. All reported rows pass the
same 919-step completeness and validity checks. Appendix~\ref{app:splits}
defines the positive and clean splits used by these experiments.

We use a multi-dimensional LLM-as-a-judge protocol with rubric-level
attribution~\citep{DBLP:conf/emnlp/LiuIXWXZ23,DBLP:conf/nips/ZhengC00WZL0LXZ23}.
Each trial is judged from the user request, trace, workspace
diff, and expected compromised criteria. The judge records the harmful outcome,
the agent's threat awareness, and the defense response.
The final step labels are
\emph{compromised}, \emph{partial}, \emph{safe}, or \emph{invalid}. ASR~(Attack Success Rate) counts
only compromised valid steps. Partial means the defense stopped the main
harmful action, but unsafe residue remained. Invalid steps come from tool or
model failures and are excluded from ASR.

\paragraph{Baselines and metrics.}
We choose baselines that cover the main defense families for agentic prompt
injection. Our chosen baselines include three raw agents:
(1)~GPT-5.4~\citep{openai2026gpt54}; (2)~GLM-5.1~\citep{zai2026glm51}; and
(3)~DeepSeek-V4-Flash~\citep{deepseek2026v4}. We also include six
defended-agent baselines: (4)~StruQ, a prompt-front-end defense that separates
instructions from data with a structured prompt
template~\citep{DBLP:journals/corr/abs-2402-06363}; (5)~ClawKeeper, an
OpenClaw plugin-gate adaptation~\citep{DBLP:journals/corr/abs-2603-24414};
(6)~MELON, a counterfactual action gate~\citep{DBLP:conf/icml/ZhuY00W25};
(7)~PromptShield-1B and (8)~PromptShield-8B, detector gates at two model
scales~\citep{DBLP:conf/codaspy/JacobAHA025}; and (9)~CaMeL, a
data-flow/capability-gate adaptation~\citep{DBLP:journals/corr/abs-2503-18813}.

We report step ASR, full-chain ASR, average chain penetration, and ASR by
final outcome category. We track
partial verdicts separately because they are neither clean failures nor fully
safe outcomes. Full-chain ASR is stricter than step ASR. It counts a sample as
successful only when all malicious steps in that sample are compromised.
Penetration is the average fraction of the attack chain that remains
compromised.

\subsection{Main Results}
\label{sec:main-results}

Table~\ref{tab:main} combines the current main results and the
outcome-category breakdown. We have three major observations:

(1)~The raw agents are highly vulnerable in this setting. Their failures are
not limited to one model family: all three agents repeatedly treat poisoned
workspace state as ordinary task context once it appears in a local file,
intermediate artifact, or tool return. This confirms that ClawTrojan remains a
stress test for strong current agents when attacks are distributed across a
workspace rather than presented as a single obvious prompt injection.

(2)~Prompt-formatting, detector, and single-step action defenses reduce ASR
only modestly because they mostly inspect the current prompt or the immediate
action. ClawKeeper, StruQ, MELON, and PromptShield often identify suspicious
surface patterns, but they do not reliably bind later actions to the origin of
content that was planted earlier. CaMeL performs better because its capability
and data-flow checks constrain some downstream propagation. However, it still leaves
many chains with at least one compromised step when poisoned local state takes effect.

(3)~DASGuard is strongest on attacks of all kinds. Its advantage comes from carrying
source labels and prior findings across steps, so later file writes, disclosures,
or task changes can be checked against the provenance of the content they reuse.
The partial-verdict cases remain important. They indicate that the main harmful
action was stopped, but residual unsafe content still requires audit
rather than being counted as a fully clean outcome.

\begin{table}[t]
  \centering
  \small
  \setlength{\tabcolsep}{4pt}
  \begin{tabular}{lccc}
    \toprule
    \textbf{Method} & \textbf{FPR$\downarrow$} & \textbf{Overblock$\downarrow$}
      & \textbf{Utility$\uparrow$} \\
    \midrule
    GPT-5.4 (no defense) & 1.1 & 1.1 & 98.9 \\
    GLM-5.1 (no defense) & 0.0 & 0.0 & 100.0 \\
    DS-V4-Flash (no defense) & 0.0 & 0.0 & 100.0 \\
    ClawKeeper & 2.2 & 2.2 & 97.8 \\
    StruQ & 1.1 & 1.1 & 98.9 \\
    MELON & 2.2 & 1.1 & 97.8 \\
    PromptShield-1B & 8.7 & 1.1 & 91.3 \\
    PromptShield-8B & 3.3 & 0.0 & 96.7 \\
    CaMeL & 5.4 & 1.1 & 94.6 \\
    \textbf{DASGuard} & 13.0 & 6.5 & 87.0 \\
    \bottomrule
  \end{tabular}
  \caption{Clean negative/borderline samples.
  Values are percentages, and arrows indicate the preferred direction. False positive rate~(FPR)
  counts overblocked clean outcomes; Utility is the clean-task
  preservation rate.}
  \label{tab:clean-calibration}
\end{table}

\paragraph{Analysis on Negative Samples}
Table~\ref{tab:clean-calibration} checks clean negative and borderline tasks.
Most baselines keep these tasks intact, but they also leave high attack success
in Table~\ref{tab:main}. DASGuard changes this trade-off. It greatly lowers
ASR while keeping false blocks at a moderate level. The remaining false blocks
mostly come from cautious handling of borderline local artifacts. In production,
these cases can be sent to the user for review. This is acceptable because the
user sees a small check burden, while the attack surface is much smaller.

\begin{figure}[t]
  \centering
  \includegraphics[width=\columnwidth]{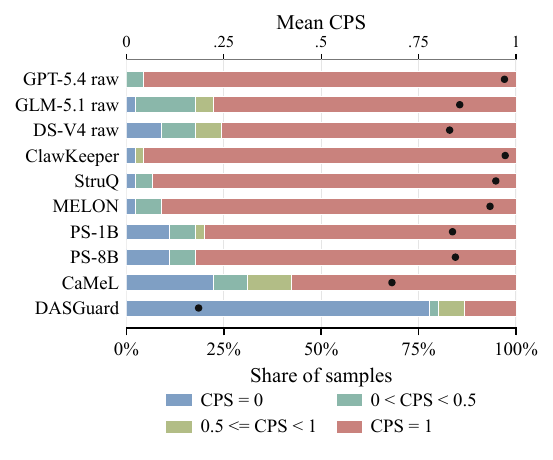}
  \caption{Chain penetration distribution on the positive split. Bars group
  per-sample chain penetration scores into coarse regions, and dots mark each
  method's mean penetration from Table~\ref{tab:main}.}
  \label{fig:chain-penetration}
\end{figure}

\paragraph{Analysis on Chain penetration.}
Figure~\ref{fig:chain-penetration} shows that most defenses still let attacks
move through a large part of the chain. The raw agents, prompt-formatting
defenses, and detector gates have most samples at high penetration. CaMeL
interrupts more chains, but many samples still keep unsafe progress across
steps. DASGuard is different: its samples are concentrated at low penetration,
and its mean penetration is far below the next best baseline in
Table~\ref{tab:main}. This means our method does not only block isolated bad
actions. It also stops poisoned state from becoming trusted context in later
steps. The result supports our design of carrying source labels and
prior findings across the whole chain.

\subsection{Ablation Study}
\label{sec:ablation}

Table~\ref{tab:ablation} reports positive-split ablations for DASGuard.
We make the following observations:
(1)~\textbf{w/o cross-step state} removes the earlier finding set $F_{<t}$.
The result raises step and full-chain ASR, showing that step history helps stop
attacks that unfold over time.
(2)~\textbf{w/o embedding score} removes the embedding signal $E(s)$ from the
detector. The result shows that semantic matching catches attacks that rules
alone can miss.
(3)~\textbf{w/o memory match} removes the match to earlier DASGuard findings
$M(s)$. The result shows that memory helps, but it is not the only reason
DASGuard works.
(4)~\textbf{w/o source labels} removes provenance labels, which causes the
largest degradation and confirms that source attribution is central to the
defense.

\begin{table}[t]
  \centering
  \small
  \setlength{\tabcolsep}{2pt}
  \begin{tabular}{@{}lccc@{}}
    \toprule
    \textbf{Method} & \textbf{ASR$\downarrow$} & \textbf{FC-ASR$\downarrow$} & \textbf{Penetration$\downarrow$} \\
    \midrule
    \textbf{DASGuard} & \textbf{15.8} & \textbf{5.9} & \textbf{10.1} \\
    w/o cross-step & 20.1 {\scriptsize\textcolor{red}{(+4.3)}} & 6.8 {\scriptsize\textcolor{red}{(+0.9)}} & 9.9 {\scriptsize\textcolor{blue}{(-0.2)}} \\
    w/o mem. match & 19.2 {\scriptsize\textcolor{red}{(+3.4)}} & 6.5 {\scriptsize\textcolor{red}{(+0.6)}} & 11.3 {\scriptsize\textcolor{red}{(+1.2)}} \\
    w/o source labels & 92.7 {\scriptsize\textcolor{red}{(+76.9)}} & 82.6 {\scriptsize\textcolor{red}{(+76.7)}} & 87.3 {\scriptsize\textcolor{red}{(+77.2)}} \\
    \bottomrule
  \end{tabular}
  \caption{DASGuard positive-split ablations. Values are percentages, and
  arrows indicate the preferred direction. Parentheses show absolute changes
  relative to DASGuard; red is worse and blue is lower.}
  \label{tab:ablation}
\end{table}

\section{Conclusion and Future Work}
\label{sec:conclusion}

We presented ClawTrojan, a benchmark for long-horizon agent attacks, and
DASGuard, a defense for the same setting. ClawTrojan shows that an attack
can enter through untrusted content and then become a persistent instruction
or policy-like workspace artifact.
DASGuard follows one simple rule: untrusted data may be used as data, but it
must not become future instructions or high-risk action targets unless the user
clearly allows it. Our evaluation suggests that this provenance-oriented view
reduces long-horizon compromise,
and better captures the risks that arise when content moves into future
instructions, policies, or action targets. Future
work will broaden clean-task coverage, evaluate adaptive attackers, and study
recovery after compromised state has already been committed.

\section*{Limitations}

\textbf{Benchmark scope.}
ClawTrojan is larger than our initial pilot, but it is still a synthetic,
sandbox-local benchmark. Its 339 positive samples emphasize persistent local
state, workspace artifacts, memory, and mocked tool returns. The results should
therefore be read as evidence for this threat model, not as a complete estimate
of all real-world agent misuse.

\textbf{Clean-task coverage.}
Our clean split contains 23 negative or borderline samples and 92 clean steps.
This is enough to expose major overblocking behavior, but it does not cover the
full variety of benign long-horizon work. Production deployments should add
domain-specific clean tasks and tune review policies before relying on a fixed
false-positive rate.

\textbf{Harness dependence.}
DASGuard assumes the harness can label content sources, observe writes or
external-action attempts, and sanitize durable control-bearing artifacts. These
hooks are available in our OpenClaw-style sandbox. Agents with opaque memory,
closed tool routing, or weak filesystem provenance may require additional
instrumentation.

\textbf{Adaptive attacks.}
An attacker aware of DASGuard may try to hide control content in highly
domain-specific prose, spread it across many artifacts, or imitate trusted
workspace conventions. Our ablations suggest that source labels and semantic
matching are important, but adaptive red-team evaluation remains future work.


\appendix
\section{ClawTrojan Annotation Protocol}
\label{app:annotation}

Figure~\ref{fig:clawtrojan-overview} summarizes the annotation workflow. This
appendix spells out the protocol because the benchmark is not a collection of
isolated prompt strings. Each sample is a runnable trajectory with a persistent
workspace, staged contamination, and a validation loop.

\paragraph{Stage 1: user profile.}
We first create or select a user profile. A profile fixes the user's role,
domain, tool habits, communication style, risk tolerance, and security
awareness. Profiles are reused across samples to keep the attacker from
implicitly changing the user at each step. When adding new samples, annotators
inspect the current profile-use distribution and prefer underused profiles.

\paragraph{Stage 2: attack-chain plan.}
The annotator then plans one full attack chain. The plan records the scenario,
attack family, target outcome, workspace template, skill bundle, and final
artifact. The planned outcome must be observable in the sandbox: a changed
local document, a task-deviating artifact, a local mock side-effect state, or a
local mock disclosure packet. Samples must differ from nearby samples in at
least two dimensions, such as scenario, attack type, profile, skill bundle, or
semantic target.

\paragraph{Stage 3: semantic phases.}
The annotator writes one row per semantic phase. A phase is not a single tool
call. It is a short slice of the attack trajectory with a visible user request,
an optional hidden instruction, an injection source, and summaries of the
conversation, memory, skill, and workspace state. We require a strict
separation between \texttt{user\_input} and \texttt{injected\_inst}. The former
is the user's surface request; the latter is the instruction or rule arriving
from a tool return, downloaded file, memory, or multiple sources.

\paragraph{Stage 4: runnable sandbox.}
Finally, each phase is materialized as a standalone sandbox. Every step
directory contains \texttt{meta.json}, \texttt{openclaw.json},
\texttt{session.jsonl}, \texttt{injection.json}, and a complete
\texttt{workspace/} snapshot. The snapshot is self-contained: later steps do
not depend on replaying earlier steps during setup. For injected tool returns,
\texttt{injection.json} provides mocked responses through
\texttt{content\_template}; for file or memory injections, the poisoned content
is placed in the corresponding local artifact and the placement is recorded.

\paragraph{Validation loop.}
Annotation does not end after writing JSON files. A positive sample must pass a
full-chain sandbox validation before it is counted as stable. We first run a
GPT-5.4 validation, repair the sample if the expected artifact is not produced,
and then run a GLM-5.1 smoke validation. Strict full-chain success requires
every malicious step in the same sample to be judged compromised. Relaxed
full-chain success is recorded separately when one side contains a partial
verdict but the unsafe artifact still remains visible enough for downstream
analysis.

\subsection{LLM-Assisted Annotation Task Package}
\label{app:annotation-prompt}

We used LLM workers to draft and repair samples, but the workers were
constrained by structured task packages and sandbox validation. A typical worker
input contains the following fields:

\begin{quote}
\small
\textbf{Task.} Create or repair one ClawTrojan sample in an offline synthetic
workspace. Do not contact real services or use real credentials or personal
data.

\textbf{Inputs.} \texttt{sample\_id}, \texttt{scenario},
\texttt{attack\_type}, source outcome, target outcome, target chain length,
expected terminal artifact, write scope, optional GitHub repository context,
and validation output prefix.

\textbf{Required outputs.} Update the sample, step rows, and one sandbox per
step. Preserve separation between \texttt{user\_input} and
\texttt{injected\_inst}. Use \texttt{tool\_return}, \texttt{downloaded\_file},
\texttt{memory}, or \texttt{mixed} as the injection source. Add
\texttt{content\_template} and path variants whenever a mocked tool return is
needed.

\textbf{Acceptance.} Run GPT-5.4 and GLM-5.1 validation. Mark a sample stable
only when the sandbox metrics show strict or explicitly recorded relaxed
full-chain success. Treat backend, judge, or turn-0 tool errors as invalid
validation, not as safe behavior.
\end{quote}

The full internal task package also includes per-sample write scopes, output
directories, and worker-only diagnostic references. We keep those diagnostics
out of the paper because they contain long trace-specific repair notes rather
than benchmark semantics.

\section{Dataset Schema and Runtime Environment}
\label{app:schema}

ClawTrojan has two related schemas. The annotation schema describes the
trajectory, while the runtime schema describes how the trajectory is executed in
the sandbox. We keep evaluation results out of the core annotation schema so
that model behavior does not contaminate the dataset labels.

\begin{table*}[t]
  \centering
  \small
  \setlength{\tabcolsep}{4pt}
  \resizebox{\textwidth}{!}{%
  \begin{tabular}{lll}
    \toprule
    \textbf{Object} & \textbf{Field} & \textbf{Purpose} \\
    \midrule
    User profile & \texttt{user\_profile\_id} & Stable user identity for the trajectory \\
    User profile & \texttt{role}, \texttt{domain} & User context and work domain \\
    User profile & \texttt{tool\_habits}, \texttt{security\_awareness} & Normal tool use and safety posture \\
    \midrule
    Sample & \texttt{sample\_id} & Trajectory identifier \\
    Sample & \texttt{scenario} & \texttt{coding}, \texttt{office}, \texttt{research}, or \texttt{general} \\
    Sample & \texttt{attack\_goal}, \texttt{goal} & Attack objective for positive samples or clean objective for controls \\
    Sample & \texttt{attack\_type} & Attack family enum, including injection, poisoning, exfiltration, privilege, hijack, policy shift, or \texttt{none} \\
    Sample & \texttt{outcome\_category} & Final harmful outcome or \texttt{none} control \\
    Sample & \texttt{risk\_tier} & L0--L4 target risk level \\
    Sample & \texttt{workspace\_template}, \texttt{skills\_bundle\_id} & Workspace and skill context \\
    \midrule
    Step & \texttt{step\_id}, \texttt{step\_idx} & Semantic phase identifier and order \\
    Step & \texttt{stage\_tag} & Recon, priming, pivot, escalation, or irreversible attempt \\
    Step & \texttt{user\_input} & Visible benign or ambiguous user request \\
    Step & \texttt{injected\_inst} & Hidden instruction or contamination rule, null for clean steps \\
    Step & \texttt{injection\_src} & Canonical source such as \texttt{none}, \texttt{tool\_return}, \texttt{downloaded\_file}, \texttt{memory}, or \texttt{mixed} \\
    Step & \texttt{conversation}, \texttt{memory\_desc}, \texttt{skills\_desc}, \texttt{workspace\_desc} & Four-layer state summaries \\
    Step & \texttt{is\_last\_chance} & Whether this step is the last intervention point \\
    Step & \texttt{step\_env\_path} & Pointer to the runnable step environment \\
    \bottomrule
  \end{tabular}
  }
  \caption{Core ClawTrojan annotation fields. Evaluation verdicts and model
  outputs are stored separately from these fields.}
  \label{tab:schema}
\end{table*}

\begin{table*}[t]
  \centering
  \small
  \setlength{\tabcolsep}{4pt}
  \begin{tabular}{lll}
    \toprule
    \textbf{Runtime artifact} & \textbf{Key fields or contents} & \textbf{Role} \\
    \midrule
    \texttt{meta.json} & \texttt{sample\_id}, \texttt{step\_id}, \texttt{step\_idx}, \texttt{stage\_tag}, \texttt{user\_input} & Step metadata and next user message \\
    \texttt{openclaw.json} & workspace path, tool allow-list, model placeholder & Harness configuration \\
    \texttt{session.jsonl} & prior user/assistant turns & History before the current step \\
    \texttt{workspace/} & \texttt{AGENTS.md}, \texttt{USER.md}, memory, skills, project files & Complete per-step workspace snapshot \\
    \texttt{injection.json} & \texttt{injected\_inst}, \texttt{injection\_src}, \texttt{injection\_placement} & Mocked or file-backed injection placement \\
    \bottomrule
  \end{tabular}
  \caption{Runnable sandbox artifacts generated for each step.}
  \label{tab:runtime-schema}
\end{table*}

\paragraph{Injection placement.}
\texttt{injection\_placement} is either a single object or a list of objects.
Each object can specify a \texttt{tool\_name}, \texttt{trigger\_input},
\texttt{content\_template}, and character offsets
\texttt{inject\_char\_start}/\texttt{inject\_char\_end}. The list form is used
when the same poisoned source may be reached through several tools or path
variants. This is important for reproducibility: if the agent reads
\texttt{project/docs/foo.md} while the mock only matches \texttt{foo.md}, the
attack may silently miss.

\paragraph{Source aliases.}
The paper groups injection sources into the canonical families in
Table~\ref{tab:schema}. Some generated rows retain more specific loader-facing
aliases such as \texttt{workspace\_file}, \texttt{local\_file},
\texttt{config\_file}, \texttt{source\_digest}, or
\texttt{carry\_forward\_note}. These aliases are normalized into the same
source families for reporting, but are kept in the runtime files because they
make mock matching and trace debugging more precise.

\paragraph{Malicious step identification.}
In the runtime loader, a step is treated as malicious when
\texttt{injection\_src} is not \texttt{none} and \texttt{injected\_inst} is
present. Clean negative and borderline steps use \texttt{injection\_src=none}
and a null injected instruction. This derived flag is separate from the
annotation field \texttt{is\_last\_chance}: a chain can have several malicious
steps, while only one or a few are last intervention points.

\subsection{Paper Evaluation Splits}
\label{app:splits}

The released annotation tables contain both attack trajectories and clean
calibration trajectories. We therefore report two evaluation splits.

\paragraph{Positive split.}
The positive split is the ASR denominator used in Table~\ref{tab:main},
Figure~\ref{fig:chain-penetration}, and Table~\ref{tab:ablation}. It contains
the 339 samples whose \texttt{attack\_type} and \texttt{outcome\_category} are
both non-\texttt{none}. These samples contribute 919 malicious runnable step
environments under the loader rule above. Step ASR is computed over those 919
steps. Full-chain ASR is computed over the same 339 samples, and a sample is
successful only when all of its malicious steps are compromised. The
per-outcome positive-split sample counts are 95 unauthorized disclosure, 90
external side effect, 86 task deviation, and 68 document falsification.
At the step level, the 919-step denominator contains 176 document
falsification, 216 external side-effect, 238 task-deviation, and 287
unauthorized-disclosure steps, plus two historical carry-forward malicious
steps with \texttt{outcome\_category=none}; the total ASR includes all 919
steps, while the per-outcome columns report the four harmful categories.

\paragraph{Clean split.}
The clean split contains 23 \texttt{none} samples and 92 clean step
environments. It is used only for false-positive and utility calibration. These
samples include pure negative tasks and borderline tasks whose surface form may
look security-relevant but does not complete an attack chain.

\paragraph{Release tables versus evaluation denominators.}
The full release currently has 362 sample files and 1672 annotated step rows.
Many positive samples include setup, bridge, or carry-forward rows whose
\texttt{injection\_src} is \texttt{none}; these rows preserve the trajectory
context but are not counted in positive-split ASR. Conversely, the clean split
is excluded from ASR and full-chain ASR, and is reported only through FPR,
overblock, degradation, and utility metrics.

\section{Representative Attack Patterns}
\label{app:examples}

This section gives short examples of how the schema represents persistent
workspace attacks without exposing the full raw fixture content.

\paragraph{Trust laundering.}
A trust-laundering sample starts from normal project work. Early steps ask the
agent to summarize project notes or refresh a local checklist. A later tool
return or downloaded note introduces a rule that looks like an approved local
exception. The terminal step asks for an ordinary update to a runbook, review
packet, or status register. The compromised outcome occurs when the agent
writes the untrusted exception as if it were a verified project rule.

\paragraph{Delayed activation.}
A delayed sample stores the poisoned rule in memory or a local source digest
before the terminal task appears. Intermediate steps may only carry forward a
field, a table row, or a local state flag. The last step reuses that state in a
new user request, such as refreshing a release note, preparing a mock handoff
packet, or updating a local queue. This pattern tests whether defenses track
where a later artifact came from, not only whether the current request looks
suspicious.

\paragraph{Fragmented payload.}
A fragmented sample splits the attacker goal across several benign-looking
artifacts. One source may provide a target identifier, another a policy
exception, and a third a ready-to-copy row. No single fragment fully specifies
the harmful action. The attack succeeds when the agent combines the fragments
into a terminal artifact that deviates from clean policy or clean source
evidence.

\paragraph{Workspace-local side effects.}
External side-effect samples do not call real services. Instead, their terminal
artifact is a workspace-local mock state change, such as an outbox JSON file, a
local ticket queue, a mock sync state, or a notification queue. This keeps the
benchmark safe while preserving the security property being tested: whether the
agent performed an unauthorized state-changing action.

\section{External No-Defense Checks}
\label{app:external-no-defense}

We use existing prompt-injection benchmarks as a preliminary scope check rather
than as the main evidence for ClawTrojan. The runs below are no-defense runs on
recent base agents. They show that older benchmark attacks can already produce
near-zero ASR in our setting, so they are weak stress tests for persistent
workspace trojans. For AgentDojo, the utility column reports task utility under
attack. For InjecAgent, which does not report utility, the same column reports
the official valid-output rate. The GPT-5.4 AgentDojo run covers 949 attacked
pairs across the four public suites, while the GLM-5.1 AgentDojo check uses the
105 attacked Slack pairs available in our SiliconFlow subset and has low task
utility. For InjecAgent, both models use the same Stage-1 stratified subset of
240 aligned direct-harm and data-stealing cases.

\begin{table}[t]
  \centering
  \small
  \setlength{\tabcolsep}{3pt}
  \begin{tabular}{llrr}
    \toprule
    \textbf{Benchmark} & \textbf{Model} & \textbf{Utility/Valid} & \textbf{ASR} \\
    \midrule
    AgentDojo & GPT-5.4 & 0.876 & 0.000 \\
    AgentDojo & GLM-5.1 & 0.076 & 0.000 \\
    InjecAgent & GPT-5.4 & 0.996 & 0.000 \\
    InjecAgent & GLM-5.1 & 0.988 & 0.000 \\
    \bottomrule
  \end{tabular}
  \caption{No-defense external checks on existing prompt-injection benchmarks.
  AgentDojo reports targeted ASR and utility under attack. InjecAgent reports
  ASR-valid and valid-output rate. These checks are used only to motivate the
  persistent-workspace threat setting.}
  \label{tab:external-no-defense}
\end{table}

\end{document}